\documentclass{emulateapj}

\shorttitle{AEGIS SEDs}
\shortauthors{Konidaris et al.}

\begin{document}

\title{AEGIS: Galaxy Spectral Energy Distributions from the X-Ray to Radio}

\author{
N.~P.~Konidaris\altaffilmark{1},
P.~Guhathakurta\altaffilmark{1},
K.~Bundy\altaffilmark{2},
A.~L.~Coil\altaffilmark{3,4},
C.~J.~Conselice\altaffilmark{5},
M.~C.~Cooper\altaffilmark{6},
P.~R.~M.~Eisenhardt\altaffilmark{7},
J.-S.~Huang\altaffilmark{8},
R.~J.~Ivison\altaffilmark{9},
S.~A.~Kassin\altaffilmark{1},
E.~N.~Kirby\altaffilmark{1},
J.~M.~Lotz\altaffilmark{10}
J.~A.~Newman\altaffilmark{3,11},
K.~G.~Noeske\altaffilmark{1},
R.~M.~Rich\altaffilmark{12},
T.~A.~Small\altaffilmark{13},
C.~N.~A.~Willmer\altaffilmark{4},
and
S.~P.~Willner\altaffilmark{8}}

\altaffiltext{1}{UCO/Lick Observatory and Department of Astronomy and
Astrophysics, University of California, Santa Cruz, CA~95064; {\tt
[npk,\,raja,\,kassin,\,ekirby,\,kai]@ucolick.org}.}

\altaffiltext{2}{California Institute of Technology, MS 105-24, 1201 East
California Boulevard, Pasadena, CA~91125; {\tt kbundy@astro.caltech.edu}.}

\altaffiltext{3}{Hubble Fellow.}

\altaffiltext{4}{Steward Observatory, University of Arizona, 933 North Cherry
Avenue, Tucson, AZ~85721; {\tt [acoil,\,cnaw]@as.arizona.edu}.}

\altaffiltext{5}{School of Physics and Astronomy, University of Nottingham,
University Park, Nottingham NG9~2RD, United Kingdom; {\tt
conselice@nottingham.ac.uk}.}

\altaffiltext{6}{Department of Astronomy, University of California at
Berkeley, 601 Campbell Hall, Berkeley, CA~94720; {\tt
cooper@astron.berkeley.edu}.}

\altaffiltext{7}{Jet Propulsion Laboratory, California Institute of
Technology, 4800 Oak Grove Drive, Pasadena, CA~91109; {\tt
prme@kromos.jpl.nasa.gov}.}

\altaffiltext{8}{Harvard-Smithsonian Center for Astrophysics, 60 Garden
Street, Cambridge, MA~02138; {\tt jhuang,\,swillner]@cfa.harvard.edu}.}

\altaffiltext{9}{Astronomy Technology Center, Royal Observatory,
Blackford Hill, Edinburgh EH9~3HJ, United Kingdom; {\tt rji@roe.ac.uk}.}

\altaffiltext{10}{National Optical Astronomical Observatories, 950 N. 
Cherry Avenue, Tucson, AZ 85719, USA; {\tt lotz@noao.edu}}

\altaffiltext{11}{Institute for Nuclear Particle Astrophysics, Lawrence
Berkeley National Laboratory, Berkeley, CA~94720; {\tt janewman@lbl.gov}.}

\altaffiltext{12}{Department of Physics and Astronomy, University of
California, Los Angeles, CA 90095; {\tt rmr@astro.ucla.edu}}

\altaffiltext{13}{Space Astrophysics, MC~405-47, California Institute of
Technology, 1201 East California Boulevard, Pasadena, CA~91125; {\tt
tas@astro.caltech.edu}.}

\setcounter{footnote}{13}

\begin{abstract}

The All-wavelength Extended Groth Strip International Survey (AEGIS)
team presents broad-band spectral energy distributions (SEDs), from
X-ray to radio wavelengths, for 71~galaxies spanning the redshift range
0.55--1.16 ($\langle${$z$}$\rangle${$\sim$}0.7).  Galaxies with secure
redshifts are selected from a small (22\,arcmin$^2$) sub-section of the
Keck\footnote{Data presented herein were obtained at the W.\ M.\ Keck
Observatory, which is operated as a scientific partnership among the
California Institute of Technology, the University of California, and NASA.
The Observatory was made possible by the generous financial support of the
W.\ M.\ Keck Foundation.}/DEIMOS galaxy redshift survey in the Extended Groth
Strip field that has also been targeted for deep panchromatic imaging by
{\it Chandra\/}\footnote{NASA's {\it Chandra\/} X-Ray Observatory was
launched in July 1999.  The {\it Chandra\/} Data Archive is part of the {\it
Chandra\/} X-Ray Center which is operated for NASA by the Smithsonian
Astrophysical Observatory.} (X-ray: 0.5--10\,keV),
GALEX\footnote{The Galaxy Evolution Explorer is a NASA Small Explorer,
launched in April~2003.  We gratefully acknowledge NASA's support for
construction, operation, and science analysis of the GALEX mission, developed
in cooperation with the Centre National d'Etudes Spatiales of France and the
Korean Ministry of Science and Technology.} (ultraviolet: 120--250\,nm),
Canada-France-Hawaii Telescope\footnote{Based on observations obtained with
the Canada-France-Hawaii Telescope which is operated by the National Research
Council of Canada, the Institut National des Science de l'Univers of the
Centre National de la Recherche Scientifique of France, and the University of
Hawaii.} (optical: 360--900\,nm),
{\it Hubble Space Telescope\/}\footnote{Based on observations with the
NASA/ESA {\it Hubble Space Telescope\/}, obtained at the Space Telescope
Science Institute, which is operated by the Association of Universities for
Research in Astronomy, Inc., under NASA contract NAS~5-26555.}
(optical/near infrared: 440--1600\,nm),
Palomar Observatory\footnote{Based on observations obtained at the Hale
Telescope, Palomar Observatory, as part of a collaborative agreement between
the California Institute of Technology, its divisions Caltech Optical
Observatories and the Jet Propulsion Laboratory (operated for NASA), and
Cornell University.} (near infrared: 1200--2200\,nm),
{\it Spitzer\/}\footnote{Based on observations made with the {\it Spitzer\/}
Space Telescope, which is operated by the Jet Propulsion Laboratory,
California Institute of Technology under a contract with NASA.  Support for
this work was provided by NASA through contract numbers 1256790, 960785, and
1255094 issued by JPL/Caltech.} (mid/far infrared: 3.6--70~$\mu$m), and the
Very Large Array\footnote{The Very Large Array of the National Radio
Astronomy Observatory is a facility of the National Science Foundation
operated under cooperative agreement by Associated Universities, Inc.}
(radio: 6--20\,cm).
The absolute magnitude of the typical galaxy in our sample is $M_B$=$-$19.82.
The ultraviolet to mid-infrared portion of their spectral energy
distributions (SEDs) are found to be bracketed by two stellar-only model
SEDs: (1)~an early burst followed by passive evolution and (2)~a constant
star-formation rate since early times; this suggests that few of these
galaxies are undergoing major starbursts.  Approximately half the galaxies
show a mid- to far-infrared excess relative to the model SEDs, consistent
with thermal emission from interstellar dust.  Two objects have power-law
SEDs, indicating that they are dominated by active galactic nuclei; both are
detected in X-rays.  The galaxies are grouped by rest-frame color,
quantitative optical morphology, and [O\,{\sc ii}] emission line strength
(possible indicator of star formation).  On average, the panchromatic SEDs of
the galaxies, from the ultraviolet to the infrared, follow expected trends:
redder SEDs are associated with red $U${$-$}$B$, early-type morphology, and
low [O\,{\sc ii}] emission, and vice versa for blue SEDs.\footnote{These data
are available at {\tt http://aegis.ucolick.org/products} .}

\end{abstract}


\section{Introduction}\label{sec:intro}  

The spectral energy distribution (SED) of a galaxy contains key clues about
the dominant physical processes responsible for energy generation.  Space-
and ground-based telescopes of the new millennium provide the sensitivity and
angular resolution to make measurements across the breadth of the
electromagnetic spectrum from X-ray to radio wavelengths.  Panchromatic SEDs
can tell us about the stellar populations within the galaxy from their
visible light [and ultraviolet (UV) and near-infrared (IR)] properties
\citep[e.g.,][and references therein---hereafter BC03]{bc03}, thermal
re-radiation of starlight in the far IR by interstellar dust \citep{dal01},
and X-ray \citep[e.g.,][]{mus80} and radio emission from active galactic
nuclei (AGNs).

\citet{dal01} have studied panchromatic SEDs of nearby galaxies and have
developed a semi-empirical model for the IR SED of normal star-forming
galaxies.  Distant galaxy SEDs are important for look-back studies of galaxy
formation and evolution.  The first such survey was GOODS \citep{gia04}.  Our
project, the All-wavelength Extended Groth Strip International Survey
\citep[AEGIS;][hereafter Paper\,I]{dav06}, extends the work of GOODS over a
larger survey area [in a part of the sky known as the Extended Groth Strip
(EGS)] thereby driving down cosmic variance.  The cornerstone of AEGIS is the
Keck/DEIMOS DEEP2 galaxy redshift survey \citep{dav03}.  Precise redshift
measurements from this survey allow us to determine the environment of each
galaxy \citep{coo05}, a key factor in determining galaxy evolution.
Moreover, these DEIMOS spectra allow us to measure linewidths and estimate
dynamical masses for the galaxies.

This paper presents SEDs of 71~optically-selected galaxies in AEGIS and is
intended to showcase the panchromatic data set.  Given the relatively coarse
angular resolution of even the state-of-the-art UV and mid-/far-IR images in
our survey, we limit our study to the {\it global\/} emission from these
distant galaxies.  The goal of this paper is to provide clean SEDs for a
handful of objects. Throughout this letter we use a cosmology of
($H_0$,\,$\Omega_{\rm m}$,\,$\Omega_\Lambda$)=(70,\,0.3,\,0.7).  Magnitudes
are in the AB system.

\section{The AEGIS Survey \& MTR Subsample}\label{sec:data} 

The AEGIS spectroscopic field is a 0.5\,deg$^2$ field at $\alpha$=14$^{\rm
h}$17$^{\rm m}$, $\delta$=+52$^\circ$30$'$.  It is only one of a few regions
in the sky with deep imaging coverage in all major observable wavebands and
spectroscopy with a 10-m class telescope (DEEP2 Keck/DEIMOS redshift survey).
Deep imaging exists in {\it Chandra\/}/ACIS (X-ray), Galaxy Evolution
Explorer (UV), {\it Hubble Space Telescope\/} ({\it HST\/}/ACS and NICMOS
(optical and near-IR), Canada-France-Hawaii Telescope (CFHT)/MegaCam and
CFH12K (multiband optical), Palomar (near-IR), {\it Spitzer\/} (mid- and
far-IR), and Very Large Array (6 and 20\,cm radio).  Availability of secure
spectroscopic redshifts is the primary sample selection criterion.  The DEEP2
survey targets galaxies with $R${$<$}24.1 and covers the redshift range
$z${$<$}1.4 in the EGS.  Detailed selection criteria and effective
wavelengths, exposure times, limiting magnitudes, source surface densities,
etc.\ for the AEGIS data sets are presented in Paper\,I.

A subfield of the EGS with complete coverage in all wavebands, known as the
Mini Test Region (MTR), is used in this letter.  The MTR consists of two
side-by-side ACS pointings (22\,arcmin$^2$) centered at $\alpha$=14$^{\rm
h}$17$^{\rm m}$52.0$^{\rm s}$, $\delta$=+52$^\circ$29$'$3.0$''$ (see Fig.\,1
of Paper\,I).  Postage stamp images around each galaxy were visually
inspected for source blending (chance superpositions or physical pairs),
cosmetic problems (CCD edge, cosmic ray, etc.), and missing entries in the
photometric catalogs for obvious sources.  Of the 101~objects with secure
redshifts in the MTR, 30 were dropped, yielding 71~objects with clean
measurements.  Ten of the dropped objects are confused/blended, 10 are
affected by cosmetic problems, and 10 are missing catalog entries [9 of the
10 are missing flux entries in all four IRAC bands because the available IRAC
photometry catalog is restricted to sources detected in band\,4 (8\,$\mu$m)].

The exclusion of the 10 blended/confused objects, some of which may be
physically interacting, does not appear to have introduced a significant bias
into our sample.  Three of the 10 cases are spectroscopically confirmed to be
chance superpositions.  No redshift information is available for the neighbor
in the remaining 7 cases, but their distribution of colors and morphologies
is identical to that of the full sample.

Our data come from a wide range of instruments, each with its own
complexities.  We have therefore relied on the expertise of each team to
obtain photometry.  The total magnitudes are derived from aperture photometry
using apertures of diameter 2--3\,times the FWHM of the PSF.  The rest-frame
absolute magnitude $M_B$ and $U${$-$}$B$ color are derived from the optical
photometry using the K-correction procedure described in \citet{wil06}.
Future papers will use more sophisticated photometry techniques involving
matched optimized apertures and PSF fitting/subtraction of neighbors.

\section{Detection Limits in the Context of Galaxy Models}\label{sec:lim} 

In this section, we examine the AEGIS detection limits in the context of
typical galaxy SEDs.  BC03 stellar population synthesis models provide a
simple conceptual framework.  These containing only stellar emission; a
complete treatment of dust extinction or emission from dust, gas, and AGN, is
beyond the scope of this work.  For simplicity, we consider two models.

The first is a passive evolution model in which the stars form early, at a
redshift $z_f$=10, over a very short period, $\sim$10\,Myr.  The stellar mass
in the model is tuned such that its $R$-band luminosity is at our detection
limit at $z$=0.7.  A galaxy with stellar mass of
$M_*$=8$\times$10$^{10}$\,$M_\odot$ is bright enough to be detected in the
AEGIS optical $R$/$I$ and {\it Spitzer\/} mid-IR bands.

Second, we consider a model in which the star-formation rate is constant from
a redshift of $z_f$=10 until it is `observed' at $z_{\rm obs}$=0.7.  As
before, the star-formation rate is tuned so that the $R$-band luminosity is
at our detection threshold at $z$=0.7.  The redshift range $z_f$=10 to
$z_{\rm obs}$=0.7 corresponds to 6.7\,Gyr.

Figure\,1 shows the SEDs of both models (red: early burst; blue: constant
star-formation rate), compared to the detection limits of our survey
(downward arrows: typically $5\sigma$ limits, see Paper\,I).  Note, the
24\,$\mu$m {\it Spitzer\/}/MIPS band's detection limit is well above the
model stellar SEDs.

\section{Spectral Energy Distributions}\label{sec:sed} 

Rest-frame SEDs for 11~galaxies are shown in Figure\,2 (left panels).  This
subsample spans a representative range of optical colors and luminosities.
The only two {\it Chandra\/} sources in our sample, most likely AGNs, are
shown at the top of the figure.  The flux density is normalized to the
$R$-band and detection limits are shown with upward arrows.  The effective
rest-frame frequency, $\nu_{\rm eff}$ of each data set is shown by a tick
mark at the top of each SED panel.  The two BC03 model SEDs described in
\S\,\ref{sec:lim} are overlaid, also normalized to the rest-frame $R$ band:
constant star-formation rate model (blue line) and early burst model (red
line).  These models do not include any dust absorption/emission.  The DEEP2
object ID, redshift, and SED type (\S\,\ref{sec:type}) are listed in the top
right corner of each left panel.  The middle panels show ``true''-color
images formed by combining the {\it HST\/}/ACS images in the $V$+$I$ bands
using the algorithm of \citet{lup03}.  The right panels show the location of
the galaxy in a plot of rest-frame absolute $B$-band magnitude versus
rest-frame $U${$-$}$B$ color (red dot) relative to other AEGIS galaxies in a
redshift slice around the target, $z_{\rm targ}${$\pm$}0.1.

Table~1 lists the SEDs of the 11~galaxies that are shown in Figure~2.  The
radio upper limits are listed in Paper\,I.

\subsection{Notes on Individual Objects}\label{sec:indiv} 

In this section, we describe the morphology, spectral features, and field
crowding of each of the 11 selected objects.  No individual galaxy in this
subsample of 11 is a VLA 6 or 20\,cm detection.  Emission lines are listed
when clearly visible by eye in order to convey a sense of the features used
for redshift determination.  For an emission line to be visible in a typical
galaxy within our sample, its equivalent width must be $\gtrsim$4\,\AA.  

12007878 is an AGN with a red core with an asymmetric underlying disk.  It
shows [O\,{\sc ii}] emission.  The {\it Chandra\/} irradiances for this
object in the hard/soft bands are:\\1.7$\pm$0.2$\times$10$^{-14}$/3.0$\pm$0.3$\times$10$^{-15}$erg\,s$^{-1}$\,cm$^{-2}$.

12007954 is an AGN, with [O\,{\sc ii}] emission and a bright nuclear source
at the center of the {\it HST\/}/ACS postage stamp.  This object is discussed
in detail by \citet{lef06}.  The {\it Chandra\/} irradiances in the hard/soft
bands are: 3.8$\pm$0.2$\times$10$^{-14}$/1.3$\pm$0.1$\times$10$^{-14}$erg\,s$^{-1}$\,cm$^{-2}$.

12011900 is a red early-type galaxy.  The bright IRAC source to the northeast
may be a low-level contaminant. 

12003965 is a faint red irregulr object, with dust lanes.  A faint optical 
companion may contaminate the IRAC fluxes.  [O\,{\sc ii}] is present.

12003798 is a spiral galaxy in the `valley' between red and blue galaxies. 
This galaxy has an [O\,{\sc iii}] doublet, [Ne\,{\sc iii}], and Balmer 
emission.

12008137 is a spiral galaxy with dust lanes.  [O\,{\sc ii}] emission is seen. 

12011707 is a dusty irregular galaxy.  Its spectrum shows [O\,{\sc ii}] and
Balmer emission.

12008269 has a very bright core and many distinct parts.  The object has
strong Balmer, [O,{\sc iii}], and [Ne,{\sc ii}].

12007899 is a blue irregular object.  The spectrum shows strong [O\,{\sc
iii}] and a Balmer emission sequence.

12012169 is a spiral galaxy, with no visible spheroidal component.  The
[O\,{\sc iii}], Balmer, and [O\,{\sc ii}] lines are strong with a hint of
weak [Ne\,{\sc iii}].

12003795 is a very blue irregular object.  It shows a sequence of strong
emission lines: [O\,{\sc ii}], [O\,{\sc iii}], Balmer lines, and the high
ionization potential lines [Ne\,{\sc iii}] and [He\,{\sc i}].

\subsection{SED Types}\label{sec:type} 

We attempt to classify each of the 71~non-blended SEDs into one of three
types.  The first type, `stellar', follow the BC03 SEDs; examples of this
class are objects 12011900, 12012169, and 12003965 in Figure\,2.  The second
type, `IR-excess', dislays a mid- to far-IR excess relative to a pure stellar
SED (IRAC 8\,$\mu$m upturn and/or MIPS 24\,$\mu$m detection); examples
include objects 12003798, 12008137, and 12008269.  The third type,
`power-law', is characterized by an SED of the form:
$S_\nu${$\propto$}$\nu^\alpha$; the only two such objects in our sample are
12007878 and 12007954.  Note, the AEGIS MIPS 24\,$\mu$m detection limit is
relatively bright (Fig.\,1) so that some of the SEDs classified as `stellar'
may well have upturns in the far-IR that fall below our detection threshold.
Finally, galaxies that do not have detections in IRAC bands\,3 or 4 are
difficult to classify; these include objects 12001707, 12007899, and
12003795.  All in all, the SED types include 22~stellar, 31~IR-excess,
2~power-law, and 16 objects that are missing IRAC fluxes.

\section{Trends with Optical Properties}\label{sec:trends} 

To give the reader an idea of the richness of the AEGIS data, each galaxy is
grouped according the following parameters: $U${$-$}$B$ color, [O\,{\sc ii}]
equivalent width, and Gini/M20 in the observer-frame $I$-band \citep*{lot04}.
The Gini coefficient and M20 combined is a quantitative measurement of the
galaxy's morphology.  These groupings are chosen for their broad-stroke
ability to create similar groups of objects, even though the parameters are
not independent.  That is, red galaxies generally have low [O\,{\sc ii}]
emission and early-type morphologies \citep{bel04,wei05} while blue galaxies
generally have strong [O\,{\sc ii}] emission and later-type morphology
\citep{ken98}.

Figure\,3 overplots all 71~SEDs (small black dots in the set of seven panels
on the left) grouped by $U${$-$}$B$, [O\,{\sc ii}] line strength, and
Gini/M20.  The grouping criteria are indicated in the top right of each
panel.  Colored dots (vertical bars) show the median normalized flux value
(range) for each band.

The set of three panels on the right show the median SEDs of each group as 
bold colored lines (where the branching indicates the `median range' as 
discussed below).  If all galaxies were detected in all bands, the median 
SED would be straightforward to compute.  In reality, not every galaxy is 
detected in all bands, so we introduce the `median range' to take 
advantage of our detection limits. To bracket the full range of 
possibilities, the upper end of the range is computed by assuming that all 
the galaxies with upper limits have fluxes that are exactly at the 
detection limit, while the lower end of the range is computed by assuming 
that they have zero flux. If all galaxies are undetected in a given band, 
then the median range is the pair (median of the normalized detection 
limits, 0).  

\section{Results}\label{sec:results} 

The near-IR to UV portions of the SEDs of the galaxies in our sample,
typically luminous galaxies at $z${$\sim$}1, are generally bracketed by the
early burst and constant star-formation rate model stellar SEDs (Fig.\,2).
This indicates that few, if any, of the luminous galaxies at these epochs are
undergoing significant starbursts.  This portion of the SED appears to be
closely correlated, at least in an average sense, with the rest-frame UV and
optical properties such as $U${$-$}$B$ color, [O\,{\sc ii}] emission line
strength, and morphology.  These same UV/optical properties do not appear to
be good predictors of whether the overall SED follows a pure stellar
population model (`stellar'), shows a mid- to far-IR excess relative to a
pure stellar population model (`IR-excess'), or follows a power law
(`power-law').  About 60\% of the galaxies have detectable mid- to far-IR
emission; this is likely thermal emission from interstellar dust.  A few
percent of the SEDs are of the form: $S_{\nu}${$\propto$}$\nu^\alpha$, likely
because they are dominated by AGNs.

\acknowledgments

The authors wish to thank D.~C.~Koo for conceiving, initiating, and helping
with this paper and the anonymous referee for valuable comments.  ALC and JAN
are supported by NASA through Hubble Fellowship grants HF-01182.01-A and
HF-01182.01-A.  The authors wish to recognize and acknowledge the very
significant cultural role and reverence that the summit of Mauna Kea has
always had within the indigenous Hawaiian community.  We are most fortunate
to have the opportunity to conduct observations from this mountain.

\clearpage

\begin{deluxetable}{ccccccccccccccc}
\setlength{\tabcolsep}{0.025in}
\tablewidth{0pt}
\tabletypesize{\footnotesize}
\tablehead{\colhead{Object} & \colhead{z\tablenotemark{a}} &
\colhead{GN\tablenotemark{b}} & \colhead{GF} & \colhead{B} &
\colhead{R} & \colhead{I} & \colhead{K} &
\colhead{IRAC1} & \colhead{IRAC2} & \colhead{IRAC3} & \colhead{IRAC4} &
\colhead{MIPS24}}
\startdata
    12007878 &     0.985 & $\downarrow   -1.551$ & $\downarrow   -1.551$ & $   -1.468$ & $   -0.726$ & $   -0.121$ & $    0.698$ & $    0.868$ & $    0.882$ & $    0.937$ & $    1.235$ & $    1.846$ & \\
    12007954 &     1.148 & $\downarrow   -1.697$ & $   -1.056$ & $   -1.144$ & $   -0.321$ & $    0.136$ & $    1.053$ & $    1.506$ & $    1.698$ & $    1.866$ & $    2.052$ & $    2.494$ & \\
    12011900 &     0.719 & $\downarrow   -0.918$ & $\downarrow   -0.918$ & $   -1.002$ & $    0.131$ & $    0.584$ & $    1.275$ & $    1.278$ & $    1.059$ & $    0.948$ & $    0.691$ & $\downarrow    1.441$ & \\
    12003965 &     0.811 & $\downarrow   -1.127$ & $\downarrow   -1.127$ & $   -1.809$ & $   -0.713$ & $   -0.197$ & $    0.478$ & $    0.463$ & $    0.294$ & $\downarrow    0.112$ & $    0.213$ & $\downarrow    1.232$ & \\
    12003798 &     0.549 & $   -1.142$ & $   -0.533$ & $   -0.394$ & $    0.399$ & $    0.672$ & $    1.275$ & $    1.210$ & $    1.063$ & $    0.935$ & $    0.931$ & $\downarrow    1.384$ & \\
    12008137 &     0.988 & $\downarrow   -1.498$ & $\downarrow   -1.498$ & $   -0.914$ & $   -0.543$ & $   -0.168$ & $    0.532$ & $    0.615$ & $    0.462$ & $    0.335$ & $    0.234$ & $    1.320$ & \\
    12011707 &     0.833 & $\downarrow   -1.318$ & $\downarrow   -1.318$ & $   -1.290$ & $   -0.822$ & $   -0.466$ & $   -0.016$ & $    0.219$ & $    0.043$ & $\downarrow   -0.079$ & $\downarrow   -0.115$ & $\downarrow    1.041$ & \\
    12008269 &     0.353 & $   -0.293$ & $    0.154$ & $    0.449$ & $    0.850$ & $    0.979$ & $    1.325$ & $    1.062$ & $    1.061$ & $    0.866$ & $    1.446$ & $    1.673$ & \\
    12007899 &     0.432 & $   -1.204$ & $   -0.873$ & $   -0.665$ & $   -0.236$ & $   -0.131$ & $    0.180$ & $   -0.143$ & $   -0.321$ & $\downarrow    0.094$ & $\downarrow    0.059$ & $\downarrow    1.214$ & \\
    12012169 &     0.845 & $\downarrow   -1.390$ & $   -0.862$ & $   -0.814$ & $   -0.675$ & $   -0.456$ & $   -0.291$ & $   -0.292$ & $   -0.408$ & $   -0.604$ & $\downarrow   -0.186$ & $\downarrow    0.969$ & \\
    12003795 &     0.755 & $\downarrow   -1.384$ & $   -1.128$ & $   -0.851$ & $   -0.623$ & $   -0.515$ & $   -0.603$ & $   -0.399$ & $   -0.605$ & $\downarrow   -0.145$ & $\downarrow   -0.180$ & $\downarrow    0.975$ & \\
\enddata
\tablenotetext{a}{Typical uncertainty in redshift is 3.4$\times$10$^{-5}$.}
\tablenotetext{b}{All flux densities are in log($\mu$Jy).  Note:
1\,$\mu$Jy=10$^{-29}$\,erg\,s$^{-1}$\,Hz$^{-1}$\,cm$^{-2}$ and AB magnitudes
are defined as $-$2.5log($I_\nu$)$-$48.6 where $I_\nu$ is in
erg\,s$^{-1}$\,Hz$^{-1}$\,cm$^{-2}$.  An upper limit is denoted by a
$\downarrow$.}
\end{deluxetable}

\clearpage

\begin{figure} 
\plotone{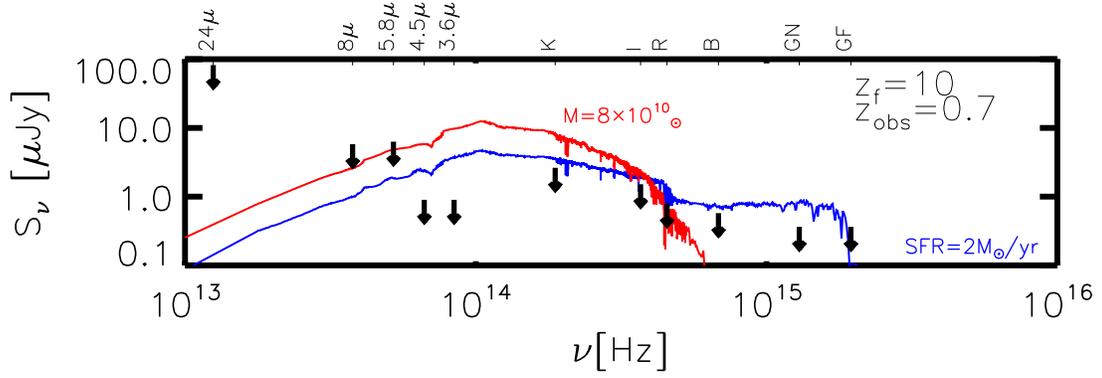} 
\caption{The AEGIS detection limits (tip of downward arrows) are compared to
BC03 stellar population models in a log flux density ($\mu$Jy) versus log
frequency (Hz) plot.  The red line represents a single early burst ($z_f$=10)
with a stellar mass of $M_*$=8$\times$10$^{10}$\,$M_\odot$ which has been
passively evolved to $z$=0.7.  The blue line represents a population that has
been forming stars at a constant rate ($\dot{M}$=2$M_\odot$\,yr$^{-1}$).  At
this rate, the galaxy forms $\sim$1.3$\times$10$^{10}$\,$M_\odot$ in stars
between $z_f$=10 and $z_{\rm obs}$=0.7.  Neither model considers dust or
nebular emission.  Model parameters are chosen to yield SEDs that are
detectable in the $R$-band.}
\end{figure}

\clearpage

\begin{figure} 
\epsscale{.55}
\plotone{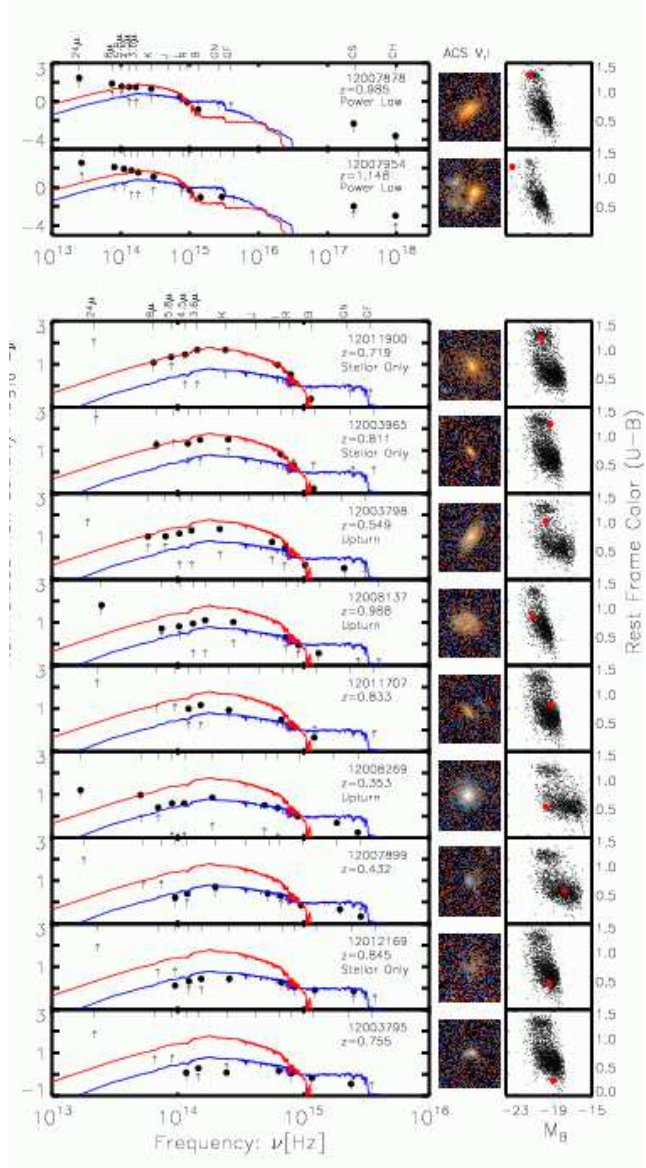}
\caption{Properties of 11~representative galaxies in the MTR that have been
hand-picked to ensure minimal neighbor contamination and optical-color
diversity.~~~
({\it Left\/})~Panchromatic SED (flux-density vs.\ frequency) of the
galaxy compared to simple stellar population models from BC03: early burst
(red) and constant star-formation rate (blue); see \S\,\ref{sec:lim} for
details.  The model SEDs tend to bracket the near-IR to UV portion of the
observed SEDs.  The DEEP2 object ID, redshift, and SED type
(\S\,\ref{sec:type}) are listed in the top right corner of each panel.
({\it Middle\/})~``True''-color image based on {\it HST\/}/ACS images in the
F606W ($V$) and F814W ($I$) bands (near right).  The images are 50\,kpc
(comoving) on a side.~~~
({\it Right\/})~Rest-frame $U${$-$}$B$ versus $M_B$ color-magnitude diagram,
showing the target galaxy (large red dot) relative to other DEEP2 galaxies
whose redshifts lie within $z_{\rm targ}${$\pm$}0.1 (small black dots).}
\end{figure}

\clearpage

\begin{figure} 
\epsscale{1.0}
\plotone{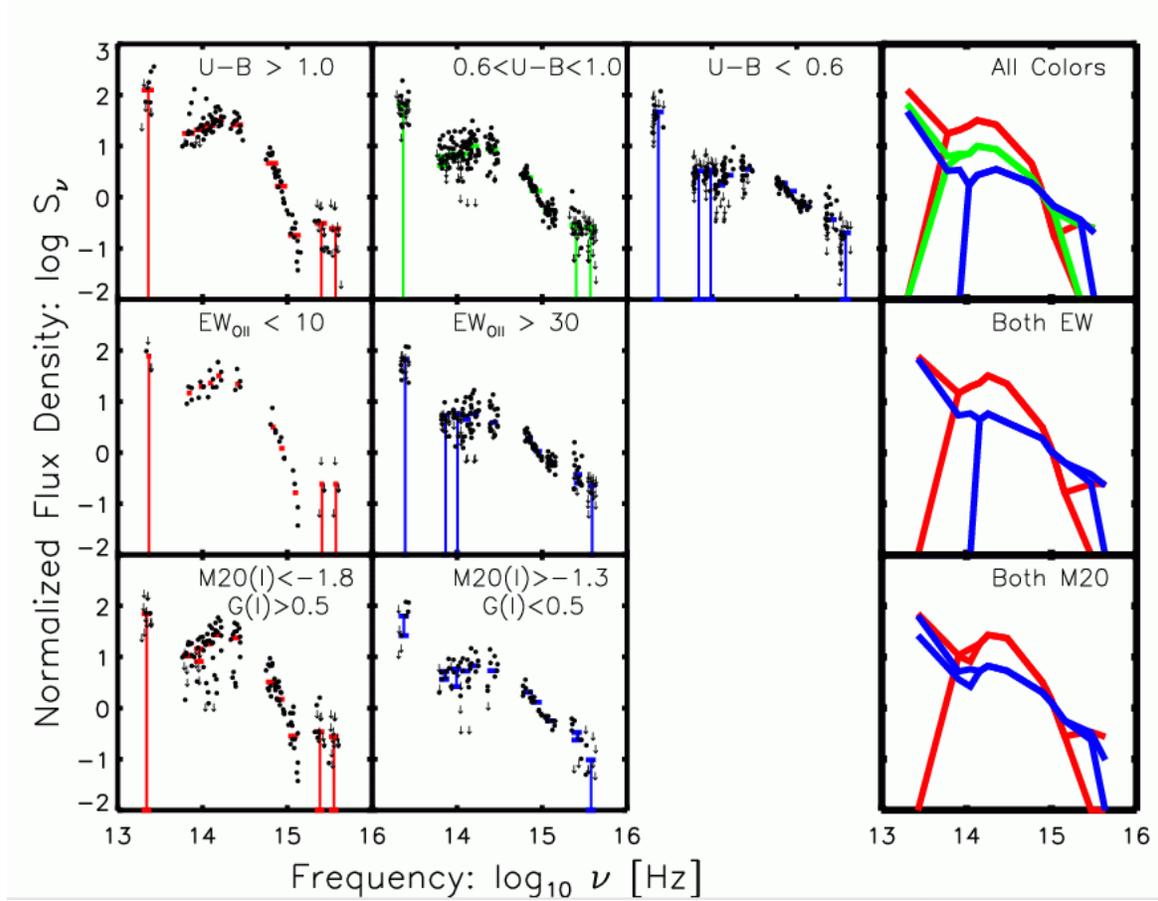}
\caption{Grouped SEDs of the galaxies in our sample in a flux density vs.\
frequency plot, with the grouping criteria indicated (set of seven~panels to
the left with thin outlines).  The small black dots indicate flux densities
for individual galaxies and the colored dots (vertical bars) indicate the
median flux density (median range) for that band (see \S\,\ref{sec:trends}).
The set of three~panels to the right with bold outlines show each group's
median SED as a bold colored line, following the same color scheme as the
panels to the left.  The median panchromatic SED shapes are seen to be
generally correlated with the optical properties on which the grouping
criteria are based---i.e.,~red rest-frame $U${$-$}$B$ color, early-type
morphology, and weak [O\,{\sc ii}] line emission are associated with SEDs
that have steep UV to IR slopes (red) while blue $U${$-$}$B$, late-type
morphology, and strong [O\,{\sc ii}] emission are associated with SEDs that
have shallow UV to IR slopes (blue).}
\end{figure}


\begin{thebibliography}{} 

\bibitem[Bell et~al.(2004)]{bel04} Bell, E.~F., et~al., \ 2004, \apj, 600,
L11 

\bibitem[Bruzual \& Charlot(2003)]{bc03} Bruzual, G., \& Charlot, S.\ 2003,
\mnras, 344, 1000

\bibitem[Cooper et~al.(2005)]{coo05} Cooper, M., Newman, J., Madgwick, D.~S.,
Gerke, B.~F., Yan, R., \& Davis, M., \ 2005, \apj, 634, 833

\bibitem[Dale et~al.(2001)]{dal01} Dale, D.~A., Helou, G., Contursi, A., 
Silbermann, N.~A., \& Kolhatkar, S.\ 2001, \apj, 549, 215

\bibitem[Davis et~al.(2003)]{dav03} Davis, M., et~al.\ 2003, Proc.\ SPIE, 
4834, 161

\bibitem[Davis et~al.(2006)]{dav06} Davis, M., et~al.\ 2006, ApJL, submitted
(this issue; astro-ph/0607355)
\notetoeditor{Davis et~al.\ 2006 is also in this AEGIS ApJL special issue.}

\bibitem[Giavalisco et~al.(2004)]{gia04} Giavalisco, M., et~al.\ 2004, \apj,
600, L93

\bibitem[Kennicutt(1998)]{ken98}Kennicutt, R.~C.\ 1998, \araa, 36, 189

\bibitem[Le~Floc'h(2006)]{lef06}Le~Floc'h, E., et~al.\ 2006, ApJL, submitted
(this issue)
\notetoeditor{Le~Floc'h et~al.\ 2006 is also in this AEGIS ApJL special issue.}

\bibitem[Lotz et~al.(2004)Lotz, Primack, \& Madau]{lot04} Lotz, J., Primack,
J., \& Madau, P.\ 2004, \aj, 128, 163

\bibitem[Lupton et~al.(2003)]{lup03} Lupton, R., Blanton, M., Fekete, G.,
Hogg, D.~W., O'Mullane, W., Szalay, A., \& Wherry, N.\ 2003, arXiv preprint
(astro-ph/0312483)

\bibitem[Mushotzky et~al.(1980)]{mus80} Mushotzky, R.~F., Marshall, F.~E.,
Boldt, E.~A., Holt, S.~S., \& Serlemitsos, P.~J.\ 1980, \apj, 235, 377

\bibitem[Weiner et~al.(2005)]{wei05} Weiner, B.~J., et~al.\ 2005, \apj, 620,
595

\bibitem[Willmer et~al.(2006)]{wil06} Willmer, C.~N.~A., et~al.\ 2006, \apj,
in press (astro-ph/0506041)

\end{thebibliography}
\end{document}